\documentclass[useAMS,usenatbib]{mn2e}
\usepackage{graphicx}
\usepackage{rotating}
\usepackage{lscape}
\usepackage{morefloats}
\DeclareMathAlphabet{\mathpzc}{OT1}{pzc}{m}{it}

\def\lsim{\,\lower2truept\hbox{${<\atop\hbox{\raise4truept\hbox{$\sim$}}}$}\,}
\def\gsim{\,\lower2truept\hbox{${> \atop\hbox{\raise4truept\hbox{$\sim$}}}$}\,}

\title[Mid-Infrared spectroscopic surveys]{Predictions for surveys with the SPICA Mid-infrared Instrument }
\author[M. Bonato et al.]
{M. Bonato$^{1,2}$\thanks{matteo.bonato@oapd.inaf.it},
M. Negrello$^{2}$,
Z.-Y. Cai$^{3}$,
G. De Zotti$^{2,4}$,
A. Bressan$^{4}$,
T. Wada$^{5}$,
\newauthor
K. Kohno$^{6,7}$,
R. Maiolino$^{8,9}$,
C. Gruppioni$^{10}$,
F. Pozzi$^{11}$
and A. Lapi$^{4,12}$\\
$^{1}$Dipartimento di Fisica e Astronomia ``G.Galilei'', Universit\`a degli Studi di Padova, Vicolo Osservatorio 3, I-35122 Padova, Italy \\
$^{2}$INAF, Osservatorio Astronomico di Padova, Vicolo Osservatorio 5, I-35122 Padova, Italy \\
$^{3}$CAS Key Laboratory for Research in Galaxies and Cosmology, Department of Astronomy, University of Science and Technology of\\ China, Hefei, Anhui 230026, China\\
$^{4}$SISSA, Via Bonomea 265, I-34136 Trieste, Italy\\
$^{5}$Institute of Space and Astronautical Science, JAXA, Yoshino-dai 3-1-1, Sagamihara, Kanagawa 229-8510, Japan \\
$^{6}$Institute of Astronomy, The University of Tokyo, Mitaka, Tokyo 181-0015, Japan \\
$^{7}$Research Center for the Early Universe (WPI), University of Tokyo, 7-3-1 Hongo, Bunkyo, Tokyo 113-0033, Japan \\
$^{8}$Cavendish Laboratory, University of Cambridge, 19 J.J. Thomson Avenue, Cambridge CB3 0HE, UK\\
$^{9}$Kavli Institute for Cosmology, University of Cambridge, Madingley Road, Cambridge, CB3 0HA, UK\\
$^{10}$INAF, Osservatorio Astronomico di Bologna, Via Ranzani 1, I-40127 Bologna, Italy \\
$^{11}$Dipartimento di Fisica e Astronomia, Universit\`a di Bologna, Via Ranzani 1, I-40127 Bologna, Italy\\
$^{12}$Dipartimento di Fisica, Universit\`a ``Tor Vergata'', Via della Ricerca Scientifica 1, I-00133 Roma, Italy}
\date{Released 2014 Xxxxx XX}

\pagerange{\pageref{firstpage}--\pageref{lastpage}} \pubyear{2014}

\def\LaTeX{L\kern-.36em\raise.3ex\hbox{a}\kern-.15em
    T\kern-.1667em\lower.7ex\hbox{E}\kern-.125emX}
\def\simlt{\mathrel{\rlap{\lower 3pt\hbox{$\sim$}}\raise 2.0pt\hbox{$<$}}}
\def\simgt{\mathrel{\rlap{\lower 3pt\hbox{$\sim$}}\raise 2.0pt\hbox{$>$}}}

\begin{document}

\label{firstpage}

\maketitle

\begin{abstract}
We present predictions for number counts and redshift distributions of galaxies detectable in continuum and in emission lines with the Mid-infrared (MIR) Instrument (SMI) proposed for the Space Infrared Telescope for Cosmology and Astrophysics (SPICA). We have considered 24 MIR fine-structure lines, four Polycyclic Aromatic Hydrocarbon (PAH) bands (at 6.2, 7.7, 8.6 and 11.3$\mu$m) and two silicate bands (in emission and in absorption) at 9.7$\mu$m and 18.0$\mu$m. Six of these lines are primarily associated with Active Galactic Nuclei (AGNs), the others with star formation. A survey with the SMI spectrometers of 1\,hour integration per field-of-view (FoV) over an area of $1\,\hbox{deg}^2$ will yield $5\,\sigma$ detections of $\simeq 140$ AGN lines and of $\simeq 5.2\times10^{4}$ star-forming galaxies, $\simeq 1.6\times10^{4}$ of which will be detected in at least two lines. The combination of a  shallow ($20.0\,\hbox{deg}^{2}$, $1.4\times10^{-1}$\,h integration per FoV) and a deep survey ($6.9\times10^{-3}\,\hbox{deg}^{2}$, $635$\,h integration time), with the SMI camera, for a total of $\sim$1000\,h, will accurately determine the MIR number counts of galaxies and of AGNs over five orders of magnitude in flux density, reaching values more than one order of magnitude fainter than the deepest \textit{Spitzer} $24\,\mu$m surveys. This will allow us to determine the cosmic star formation rate (SFR) function down to SFRs more than 100 times fainter than reached by the \textit{Herschel} Observatory.
\end{abstract}


%%%%%%%%%%%%
% FIGURE 1 %
%%%%%%%%%%%%
\begin{figure*}
  \hspace{+0.0cm} \makebox[\textwidth][c]{
    \includegraphics[trim=1.3cm 1.6cm 1.7cm
    2.0cm,clip=true,width=1.0\textwidth,
    angle=0]{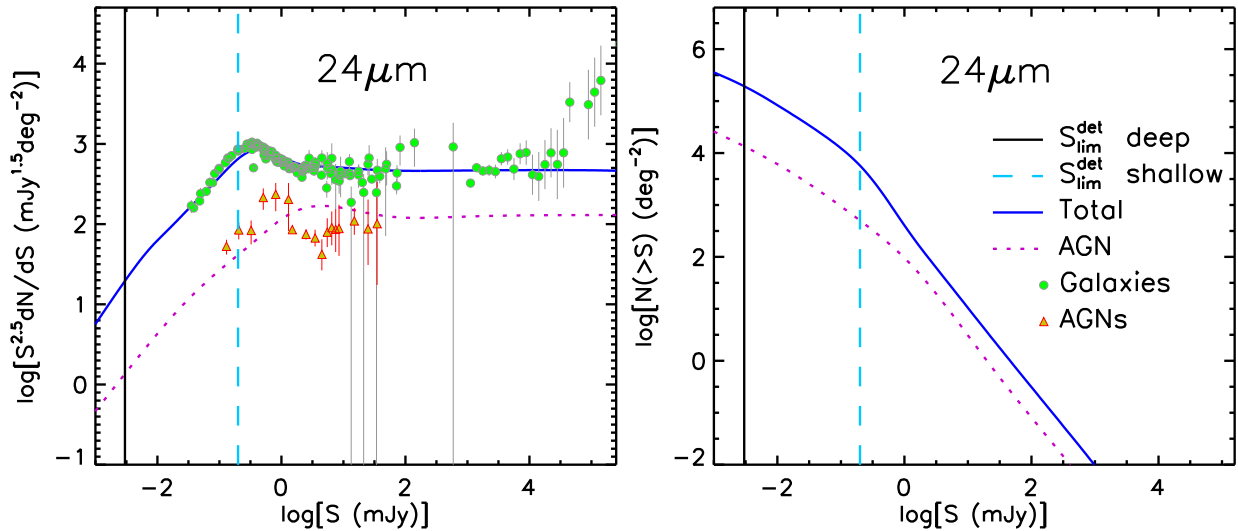}
  }
  % \vspace{-4.5cm}
  \caption{Euclidean normalized differential number counts (on the left) and integral counts (on the right) at 24\,$\mu$m given by the adopted model for galaxies as a whole (starburst plus AGN components; solid blue lines) and for AGNs alone (dotted violet lines). The vertical solid black and dashed cyan lines correspond to the SMI camera detection limits for the proposed deep and shallow surveys, respectively.  Counts of galaxies (filled green circles) are from \citet{Takagi12}, \citet{Clements11}, \citet{Bethermin2010}, \citet{LeFloch09}, \citet{Shupe08}, \citet{Papovich04}   and \citet{Sanders2003}. The latter data show an excess at the highest flux densities due to the enhanced number of bright galaxies in the Virgo supercluster.  AGN counts (red triangles) are from \citet{Treister06} and \citet{Brown06}.}
  \label{fig:intcounts_photometry}
\end{figure*}

%%%%%%%%%%%%%%%%%%
\section{Introduction}\label{sect:intro}
%%%%%%%%%%%%%%%%%%

Studying the co-evolution of star-formation and black-hole accretion is one of main scientific goals of the SPace InfraRed telescope for Cosmology and Astrophysics (SPICA)\footnote{http://www.ir.isas.jaxa.jp/SPICA/SPICA\_HP/index-en.html}. SPICA will be equipped with two main instruments: the SpicA FAR infrared Instrument \citep[SAFARI;][]{Roelfsema12} and the SPICA Mid-infrared Instrument (SMI)\footnote{https://home.sron.nl/files/LEA/SAFARI/spica\_workshop\_2014 /SMI\_factsheet2.pdf \\
https://home.sron.nl/files/LEA/SAFARI/spica\_workshop\_2014 /KanedaH\_SPICA\_workshop\_2014.pdf}. In \cite{Bonato2014a,Bonato2014b} we have presented detailed predictions for the number counts and the redshift distributions of galaxies detectable in blind spectroscopic surveys with SAFARI, accounting for both the starburst and the Active Galactic Nucleus (AGN) components. Here we focus on the SMI.

The SMI has two basic observing modes: the wide-field imaging camera mode and the spectrometer mode with two detectors \citep[Spec-S and Spec-L;][]{Kataza12}. The technical specifications for the SMI used in this work are: R=1000 spectrometers, $\hbox{FoV}=150''\times3''$, wavelength ranges $20-27\,\mu$m (Spec-S) and $27-37\,\mu$m (Spec-L); R=20 wide field camera, $\hbox{FoV}=5'\times5'$, wavelength range $20 -37\,\mu$m. The spatial resolution (FWHM) varies from $1.4^{\prime\prime}$ at $20\,\mu$m to $2.6^{\prime\prime}$ at $37\,\mu$m. The line detection limit (1\,hr, $5\,\sigma$) is in the range $6-23\,\times 10^{-20}\,\hbox{W\,m}^{-2}$ for the camera, $2-7\,\times 10^{-20}\,\hbox{W\,m}^{-2}$ for the Spec-S and $3-10\,\times 10^{-20}\,\hbox{W\,m}^{-2}$ for the Spec-L. The point source continuum sensitivity (1\,hr, $5\,\sigma$) in a low background region increases from $\sim10\,\mu$Jy at $20\,\mu$m to $\sim60\,\mu$Jy at $37\,\mu$m; at $24\,\mu$m it is $\simeq 10.5\,\mu$Jy. The survey speed for the $5\,\sigma$ detection of a point source with a continuum flux density of $40\,\mu$Jy with the camera is $\sim 7\,\hbox{arcmin}^2$/hr; for the detection of the line flux of $2\times 10^{-19}\,\hbox{W\,m}^{-2}$ it is $\sim 4\,\hbox{arcmin}^2$/h for Spec-S and $\sim 2\,\hbox{arcmin}^2$/h for Spec-L.

% In previous works (\citealt{Bonato2014a,Bonato2014b}) we presented
% predictions for the source counts detectable by SpicA FAR infrared
% Instrument \citep[SAFARI;][]{Roelfsema12} spectroscopic surveys.
The SMI instrument is crucial to enhance the outcomes of the spectroscopic surveys carried out with SAFARI. The $R=1000$ spectrometer is needed to observe fine structure lines with a resolution similar to SAFARI, whereas the $R=20$ wide field camera is essential to uncover star-forming galaxies in the four broad and very bright Polycyclic Aromatic Hydrocarbon (PAH) bands at 6.2, 7.7, 8.6 and 11.3$\mu$m. The lines that will be detected can come either from star forming regions or from nuclear activity or from both.

The SMI spectroscopy will allow us to exploit the rich suite of MIR diagnostic lines to trace the star formation and the accretion onto the super-massive black holes up to high redshifts through both blind spectroscopic surveys and pointed observations. The MIR lines detectable by the SMI provide excellent diagnostics of the gas density and of the hardness of the exciting radiation field. The ratios of two lines having similar critical density and different ionization potential allow us to estimate the ionization of the gas, while the ratios of two lines with different critical density and similar ionization  potential provide estimates of the gas density in the region \citep{Spin92}. A comprehensive discussion of infrared density indicators is given by \citet{Rubin89}. As shown in \citet{Sturm02}, mid-IR line ratio diagrams can be used to identify composite sources and to distinguish between emission from star forming regions  and emission excited by nuclear activity. These diagnostic diagrams are constructed by plotting pairs of line ratios against each other (for example [NeVI]\-7.63/[OIV]\-25.89 and [NeVI]\-7.63/[NeII]\-12.81), in which different types of regions can be easily separated and distinguished. Similar diagnostic diagrams with different sets of weaker lines have been proposed by \citet{Spin92} and \citet{Voit92}. The multiplicity of possible combinations of lines allows us to adapt these diagnostic tools to different redshift ranges. In this respect, the complementary wavelength coverages of SMI and SAFARI substantially enhances the potential of the SPICA mission.

The SMI camera will also substantially improve our knowledge of source counts in the mid-infrared (MIR) region by extending them to much fainter flux density levels than achieved by \textit{Spitzer} and reaching a much better statistics. This allows a considerable improvement of our understanding of the cosmic star formation history.

%\textbf{Compared to the two previous papers \cite{Bonato2014a,Bonato2014b} devoted to predictions for future blind IR spectroscopic surveys (performed especially by SPICA/SAFARI), the present paper deals with some new aspects. They include an assessment of empirical or theoretical (where observational data were insufficient) correlations between line and continuum luminosities for other 23 MIR/FIR lines, and predictions for number counts and redshift distributions of galaxies detectable in emission lines and also in continuum with the SPICA/SMI instrument.}

In this paper we use the \citet{Cai13} evolutionary model as upgraded by \citet{Bonato2014b}. The model deals in a self consistent way with the emission of galaxies as a whole, including both the starburst and the AGN component, and was successfully tested against a large amount of observational data.

The plan of the paper is the following. In Section\,\ref{sect:evol} we briefly summarize the adopted model for the evolution with cosmic time of the IR (8-1000\,$\mu$m) luminosity function. In Section\,\ref{sect:imaging} we discuss imaging observations with the wide field SMI camera. In Section\,\ref{sect:line_vs_IR} we present the relations between line and continuum luminosity for the main MIR lines. In Section\,\ref{sect:LF} we work out our predictions for line luminosity functions, number counts and redshift distributions within the SMI wavelength coverage. In Section\,\ref{sect:survey} we discuss possible SMI observational strategies. Section\,\ref{sect:concl} contains a summary of our main conclusions.

We adopt a flat $\Lambda \rm CDM$ cosmology with matter density $\Omega_{\rm m} = 0.32$, dark energy density $\Omega_{\Lambda} = 0.68$ and Hubble constant $h=H_0/100\, \rm km\,s^{-1}\,Mpc^{-1} = 0.67$ \citep{PlanckCollaborationXVI2013}.

%%%%%%%%%%%%
% FIGURE 2%
%%%%%%%%%%%%
\begin{figure}
  \hspace{+0.0cm} %\makebox[\textwidth][c]{
   \includegraphics[trim=2.8cm 0.6cm 1.0cm  0.5cm,clip=true,width=0.45\textwidth,    angle=0]{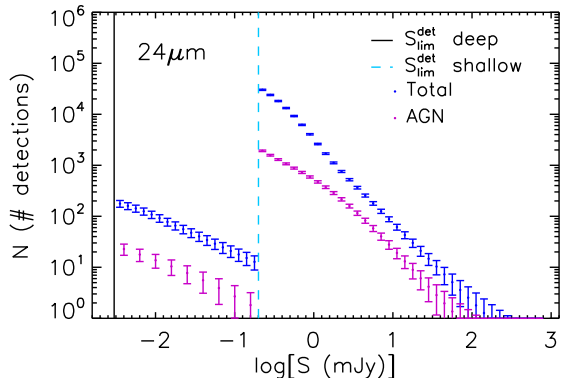}
%    \includegraphics[width=0.55\textwidth,angle=0]{detections_wedding_cake_1_24micron.eps}
  %}
  % \vspace{-4.5cm}
  \caption{Number of predicted detections of galaxies as a whole, including the AGN component (upper blue points) and AGNs alone (lower violet points) within $\Delta\log S=0.1$ bins ($\Delta\log S=0.2$ bins for faint AGNs) for the proposed shallow and deep surveys with the SMI camera. The error bars are the sum in quadrature of Poisson errors and of the variance due to clustering (sampling variance; see text). The latter is relevant only for the deep survey. The vertical solid black and dashed cyan lines correspond to the survey $5\,\sigma$ detection limits.}
  \label{fig:detections}
\end{figure}

%%%%%%%%%%%%
% FIGURE %
%%%%%%%%%%%%
%\begin{figure*}
%  \hspace{+0.0cm} \makebox[\textwidth][c]{
%    \includegraphics[trim=0.2cm 0.7cm 1.2cm
%    0.9cm,clip=true,width=1.0\textwidth,
%    angle=0]{flux_ratios_photometry.eps}
%  }
  % \vspace{-4.5cm}
%  \caption{Ratios of flux densities at the central wavelengths of adjacent SMI camera filters for the three SEDs used for the model calculations: starburst galaxy (solid black lines), proto-spheroidal galaxy (dotted red lines) and type 2 AGN (dashed blue lines).}
%  \label{fig:diff_pop_filters}
%\end{figure*}

%%%%%%%%%%%%%%%%%%%%%%%%%%%%%%%%%%%%%%%%%%%
\section{Evolution of the IR luminosity functions}\label{sect:evol}

The \cite{Cai13} model, adopted here, is based on a comprehensive ``hybrid'' approach that combines a physical model for the progenitors of the spheroidal components of galaxies (early-type's and massive bulges of late-type's) with a phenomenological one for disk components. The evolution of the former population  is described by an updated version of the physical model by \citet[][see also Lapi et al. 2006, 2011]{Granato2004}. In the local universe these objects are composed of relatively old stellar populations with mass-weighted ages $\gsim 8$--9\,Gyr, corresponding to formation redshifts $z\gsim 1$--1.5, while the disc components of spirals and the irregular galaxies are characterized by significantly younger stellar populations \citep[cf.][their Fig.\,10]{Bernardi2010}. Thus the progenitors of spheroidal galactic components, referred to as proto-spheroidal galaxies or ``proto-spheroids'', are the dominant star forming population at $z\gsim 1.5$, while IR galaxies at $z\lsim 1.5$ are mostly galaxy disks.
% \citet{Cai13} have worked out an analytic
% formula for the probability distribution of the starburst or of the
% AGN luminosity of a proto-spheroid given its total luminosity
% \citep[see also][]{Bonato2014b}.

In the case of proto-spheroids, the model describes the co-evolution of the stellar and of the AGN component, allowing us to deal straightforwardly with objects as a whole. This does not happen for disk components of late-type galaxies, whose evolution is described by a phenomenological, parametric model, distinguishing between the two sub-populations of ``cold'' (normal) and ``warm'' (starburst) galaxies. AGNs are treated as a separate population. Following \cite{Bonato2014b}, we have associated them to the late-type galaxy populations using the \citet{Chen13} correlation between star formation rate (SFR) and black hole accretion rate. Both type 1 and type 2 AGNs, with relative abundances, as a function of luminosity, derived by \citet{Hasinger08} \citep[see][for a review]{Bianchi12} are taken into account.

The \citet{Chen13} correlation does not apply to bright optically selected QSOs, which have high accretion rates but are hosted by galaxies with SFRs ranging from very low to moderate. As in \citet{Bonato2014b} we reckon with these objects adopting the best fit evolutionary model by \citet{Croom2009} up to $z=2$. As shown by \citet{Bonato2014b}, this approach reproduces the observationally determined bolometric luminosity functions of AGNs at different redshifts. At higher redshifts optical AGNs are already accounted for by the \citet{Cai13} model which also accounts for redshift-dependent AGN bolometric luminosity functions.

%%%%%%%%%%%%
% FIGURE 3%
%%%%%%%%%%%%
\begin{figure*}
  \hspace{+0.0cm} \makebox[\textwidth][c]{
    % \includegraphics[width=1.55\textwidth,
    % angle=0]{../LFIR/Grafici_articolo/panel_intcounts.ps}
    \includegraphics[trim=0.6cm 2.6cm 1.4cm
    3.2cm,clip=true,width=1.0\textwidth,
    angle=0]{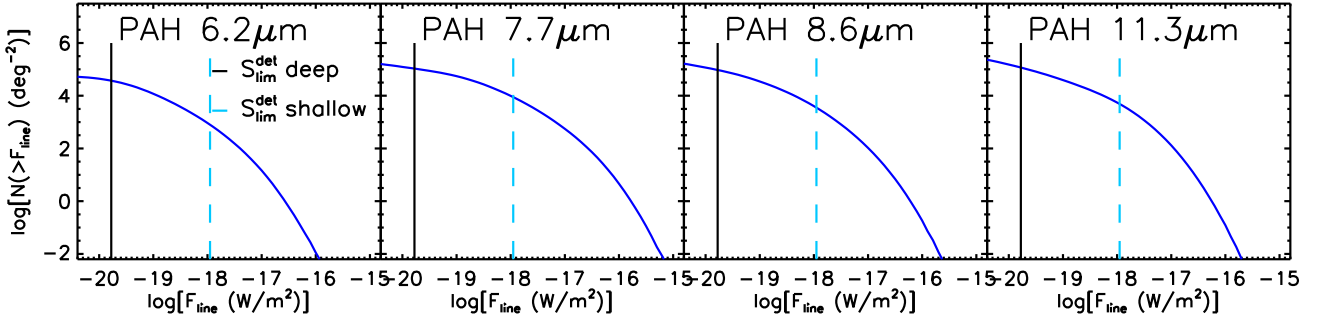}
  }
  % \vspace{-4.5cm}
  \caption{Integral counts in 4 PAH lines of star-forming galaxies over the SMI wide field camera wavelength range. The solid black and the dashed cyan vertical lines correspond to the SMI camera detection limits for the proposed deep and shallow surveys, respectively.}
  \label{fig:intcounts_spica_all_WFC}
\end{figure*}

%%%%%%%%%%%%
% FIGURE 4%
%%%%%%%%%%%%
\begin{figure*}
  \hspace{+0.0cm} %\makebox[\textwidth][c]{
    % \includegraphics[width=1.55\textwidth,
    % angle=0]{../LFIR/Grafici_articolo/panel_intcounts.ps}
%\includegraphics*[trim=8cm 0.6cm 5cm  0.5cm,clip=true,width=0.3\textwidth, angle=0]{z_distr_tot_hybrid_WFC_shallow.eps}
%\includegraphics*[trim=8cm 0.6cm 5cm  0.5cm,clip=true,width=0.3\textwidth, angle=0]{z_distr_tot_hybrid_WFC_deep.eps}
\includegraphics*[trim=1.6cm 0.1cm 1.6cm  0.1cm,clip=true,width=0.45\textwidth, angle=0]{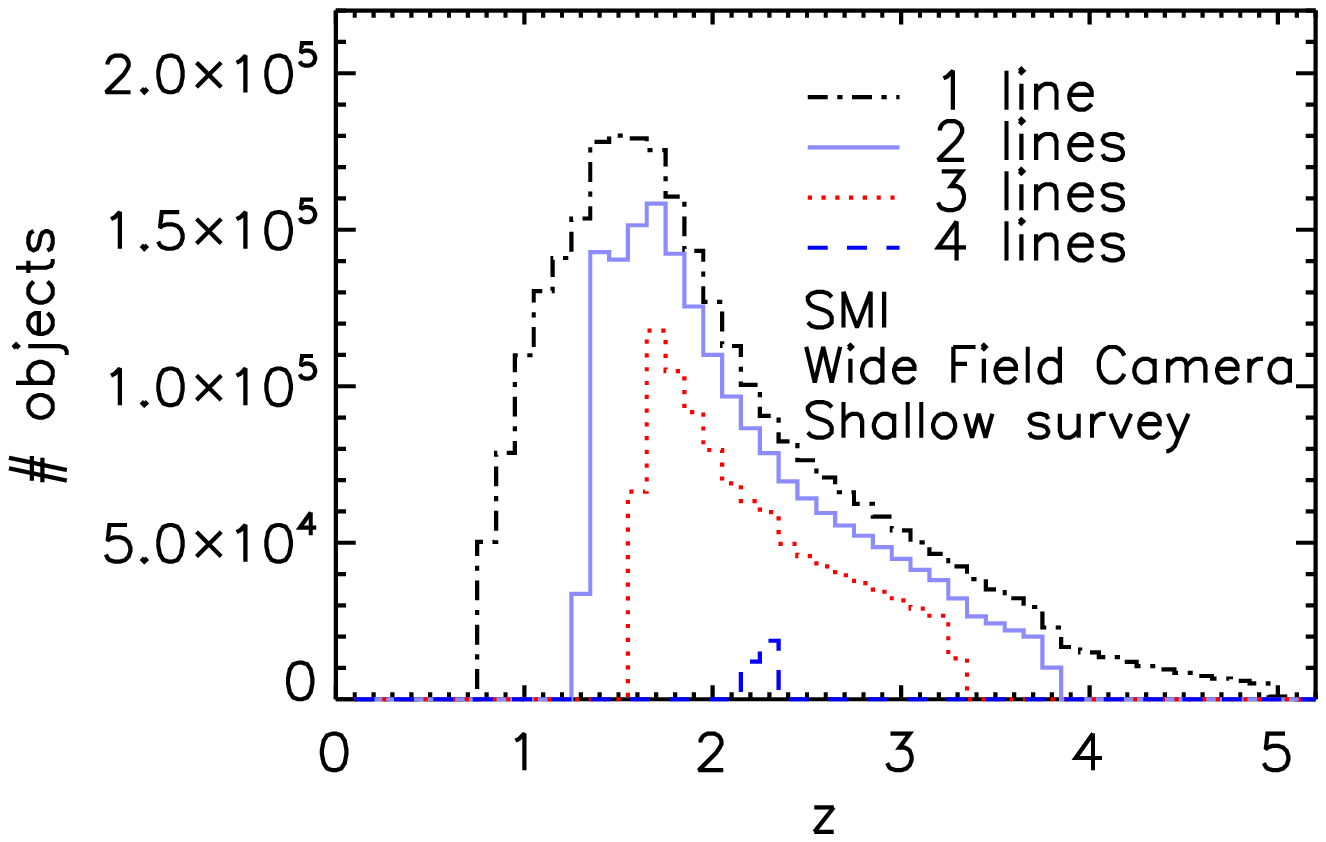}
\includegraphics*[trim=1.6cm 0.1cm 1.6cm  0.1cm,clip=true,width=0.45\textwidth, angle=0]{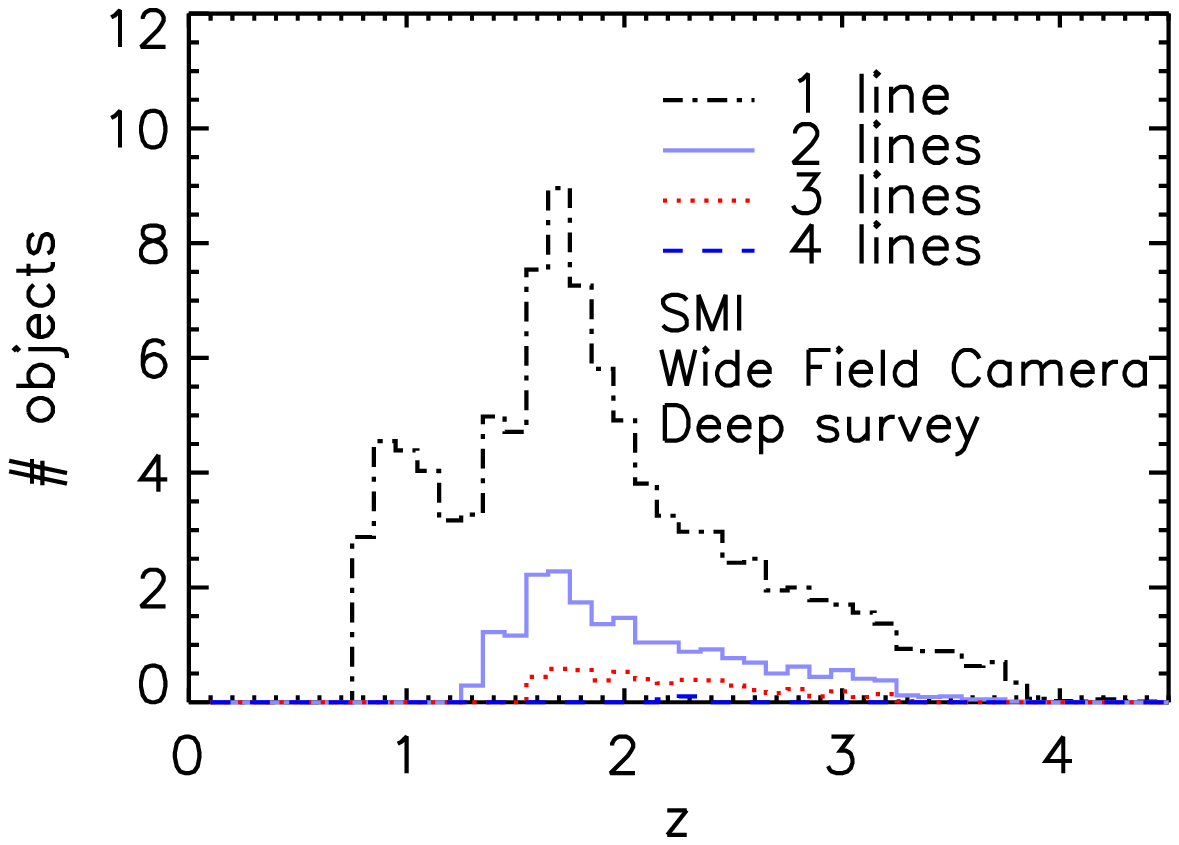}

%  }
  % \vspace{-4.5cm}
  \caption{Redshift distributions of galaxies detected in at least 1 to 4 PAH lines by the proposed shallow (left panel) and deep (right panel) survey with the SMI wide-field camera.}
  \label{fig:zdistr_WFC}
\end{figure*}

%%%%%%%%%%%%%%%%%%%%%%%%%%%%%%%%%%%%%%%%%%%
\section{Surveys with the wide field camera}\label{sect:imaging}

The SMI camera can substantially extend the mid-IR flux density range observed so far (by the \textit{Spitzer} satellite). Using the counts yielded by the model, that are strongly constrained by observational data (see the left panel of Fig.~\ref{fig:intcounts_photometry}), we find a $24\,\mu$m $5\,\sigma$ confusion limit of $4.8\times10^{-5}\,$mJy. For comparison, the deepest \textit{Spitzer} surveys reached $30\,\mu$Jy using an extraction technique based on prior source positions at $3.6\,\mu$m \citep{Magnelli2011}; the completeness of the resulting catalog, estimated via Monte Carlo simulations to be 80\%, is however difficult to assess. As illustrated by Fig.~5 of \citet{Bethermin2010} at fluxes lower than $8.0\times10^{-2}\,$mJy the $24\,\mu$m counts  are endowed with substantial uncertainties due either to poor statistics or to substantial corrections for incompleteness.

With the current technical specifications (see Sect.~\ref{sect:intro}), reaching a $5\,\sigma$ limit of $3.0\,\mu$Jy, i.e. going at least one order of magnitude deeper than \textit{Spitzer}, requires $\simeq 635\,$h per FoV. Observations of a single FoV ($6.9\times10^{-3}\,\hbox{deg}^{2}$) will be enough to detect hundreds of sources per 0.1 dex in flux density. A survey at this limit would resolve $\simeq 92\%$ of the background estimated using the adopted model. A determination of the source counts over a very broad flux density interval can be achieved adding a shallow survey with a detection limit of $0.2\,$mJy, an integration time of $\sim1.4\times10^{-1}$\,h per FoV, covering an area of $20.0\,\hbox{deg}^{2}$. The global observing time (without overheads) is then $1046\,$h.

Not surprisingly, the deep survey would be very costly of time. Is there a sufficient scientific motivation for it, given that the $24\,\mu$m counts are found to be rapidly converging already at much brighter flux density levels? The simplest answer is that fully exploiting the potential of a new instrument that provides a large boost in sensitivity is a must since it is the most direct way for exploring the unknown, looking for the unexpected. On top of that, in this case going deeper guarantees important information on a still poorly understood aspect of galaxy evolution: the effect of feedback on the star formation history of low mass galaxies at high redshifts, hence also on the build up of larger galaxies via mergers.

It is generally agreed that the flatter slope of the faint end of the galaxy luminosity function compared to that of the halo mass function is due feedback, but how the feedback operates  is still unclear \citep[e.g.,][]{SilkMamon2012}. Various flavors of feedback have been advocated, including re-ionization, supernova  explosions, tidal stripping and more, but the role of each effect is not well understood. The high surface density reached by the deep survey (right-hand panel of Fig.~\ref{fig:intcounts_photometry}) implies that it will detect low mass galaxies, with low SFRs at substantial redshifts, as discussed at the end of this section. Thus it will provide direct information on a key aspect of galaxy evolution.

%\textbf{Alternatively, if we would be satisfied of a deep survey with a detection limit two times greater, the observational time required by this deep survey would be only $\simeq 159\,$h per FoV and the fraction of resolved background would be $\simeq 88\%$.}

The right-hand panel of Fig.~\ref{fig:intcounts_photometry} shows the predicted integral counts at 24\,$\mu$m of galaxies as a whole (starburst plus AGN component; solid blue lines) and of the AGNs alone (dotted violet lines). The proposed surveys are expected to detect $\simeq 1.1\times10^{5}$ galaxies and $\simeq 1.0\times10^{4}$ AGNs (shallow survey), and $\simeq 1.3\times10^{3}$ galaxies and $\simeq 94$ AGNs (deep survey). Interestingly, the surface density of AGNs detected by the deep survey is essentially the same as that of the deepest Chandra survey \citep[4 million seconds;][]{BrandtAlexander2015} in X-rays. The mid-IR AGN counts are crucial, among other things, to assess the abundance of heavily absorbed AGNs, missed by X-ray surveys, that can contribute an important fraction of the high-energy X-ray background.

Figure~\ref{fig:detections} shows the number of predicted detections of galaxies and AGNs in $\Delta \log S=0.1$ bins ($\Delta\log S=0.2$ for AGNs in the case of the deep survey). Below 0.2\,mJy, where observational data are largely absent (see Fig.~\ref{fig:intcounts_photometry}), the proposed survey will detect from $\sim13$ to $\sim177$ galaxies per $\Delta\log S=0.1$ bin, and from $\sim2$ to $\sim23$ AGNs per $\Delta\log S=0.2$ bin. At 10\,mJy, where the \textit{Spitzer} counts have a very poor statistics (Fig.~\ref{fig:intcounts_photometry}), the proposed survey will detect $\sim76$ galaxies and $\sim24$ AGNs per $\Delta\log S=0.1$ bin.

The error bars plotted in Fig.~\ref{fig:detections} include both the Poisson fluctuations and the contribution from the sampling variance. The latter is due to the field-to-field variations arising from source clustering and is important especially in the case of surveys covering small areas. The total fractional variance of the differential counts, $n$, can be written as \citep{Peebles1980}:
\begin{equation}
\left\langle {n-\langle n \rangle \over \langle n\rangle}\right\rangle^2 = {1\over \langle n \rangle} + \sigma_{\rm cl}^2
\end{equation}
with
\begin{equation}\label{eq:sigmav}
\sigma_{\rm cl}^2={1\over \Omega^2}\int \int w(\theta)\,d\Omega_1\,d\Omega_2
\end{equation}
where $\langle n \rangle$ is the mean count in the flux density bin, $\theta$ is the angle between the solid angle elements $d\Omega_1$ and $d\Omega_2$, $w(\theta)$ is the angular correlation function and the integrals are over the solid angle, $\Omega$, covered by the survey.

The angular correlation function of the faint sources detectable by the SMI camera is not known.  \citet{Fang2008} obtained an angular clustering amplitude $A\simeq 0.001$ at one degree for the galaxies detected at $24\,\mu$m by the \textit{Spitzer} Wide-Area Infrared Extragalactic Survey (SWIRE) with $S_{24\mu\rm m}\simgt 350\,\mu$Jy. The fainter galaxies detected by the SMI camera are presumably less luminous on average and therefore are unlikely to have a higher clustering amplitude. The slope of the $24\,\mu$m angular correlation function is uncertain. Adopting the standard value of $-0.8$ (i.e. $w(\theta) \simeq 0.001 (\theta/\hbox{deg})^{-0.8}$) we get \citep{DeZotti2010}:
\begin{equation}
\sigma_{\rm cl}^2=2.36\times 10^{-3} (\Omega/\hbox{deg}^2)^{-0.4}.
\end{equation}
In the case of a survey over a single FoV $\sigma_{\rm cl}^2=0.017$. This is a significant, although minor,  contribution to the error budget for the proposed deep survey. For the shallow survey, the contribution due to the sampling variance is negligible.

A big plus of the SMI camera is its full coverage of the 20-37\,$\mu$m spectral range with $R=20$ resolution. Using the PAH emission template described in \citet[][their eq. (7)]{Groves08}  and the normalized parameters of the Lorentzian components of the PAH emission band given  in their Table\,2 we have verified that the four PAH bands we consider fill the spectral resolution element of the camera for the whole redshift range of interest, so that their signal is not diluted. In the worst case the fraction of the PAH line flux falling within a resolution element varies from $\sim 0.8$ (PAH $7.7$ and $8.6\,\mu$m) to $\sim 1$ (PAH $6.2$ and $11.3\,\mu$m).

Coupling the relationships between the PAH and the IR luminosities discussed in Sect.~\ref{sect:line_vs_IR} with the redshift dependent IR luminosity functions given by the model we have computed the integral counts of galaxies in the four PAH lines over the wavelength range covered by the SMI wide field camera (see Fig.~\ref{fig:intcounts_spica_all_WFC}). %Based on the PAH emission template described in \citet{Groves08} [their eq. (7)] and the normalized parameters of the Lorentzian components of the PAH emission band given  in their Table\,2 we have concluded that, over the full redshift range relevant here, the limited spectral resolution of the camera does not cause a significant dilution of the PAH features. %verified that the four PAH lines we consider fill the spectral resolution element of the camera ($R=20$), so that their signal is not diluted. The fraction of the PAH line flux falling within a resolution element varies from $\sim 0.8$ (PAH $7.7$ and $8.6\,\mu$m) to $\sim 1$ (PAH $6.2$ and $11.3\,\mu$m).
We find that the proposed shallow survey will detect $\simeq 3.0\times10^{6}$ galaxies in at least one PAH line and $\simeq 1.9\times10^{6}$ in at least two lines; for the deep survey the number of detections are $\simeq 100$ in at least one line and $\simeq 20$ in at least two lines. The redshift distributions of galaxies detected in 1, 2, 3 and 4 lines are shown in Fig.~\ref{fig:zdistr_WFC}.

\begin{figure} % Fig. 5
  \hspace{+0.0cm} %\makebox[\textwidth][c]{
    \includegraphics*[trim=2.8cm 0.6cm 0.7cm 0.8cm,clip=true,width=0.5\textwidth, angle=0]{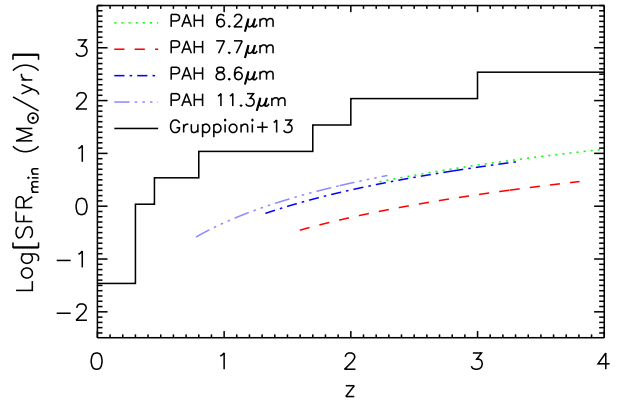}
%    \includegraphics*[width=0.5\textwidth, angle=0]{SFR_min.eps}

  %}
  % \vspace{-4.5cm}
  \caption{Comparison between the minimum SFR achieved by the proposed deep survey with the SMI wide field camera (through spectroscopic detections of PAH lines) and the SFR corresponding to the minimum luminosities represented in the IR luminosity functions determined by \citet{Gruppioni13} on the basis of Herschel/PACS and SPIRE surveys, as a function of the redshift.}
  \label{fig:SFR}
\end{figure}

Figure~\ref{fig:SFR} shows the minimum SFR \citep[calculated using our line/$L_{\rm IR}$ relations and the $L_{\rm IR}$/SFR relation by][]{KennicuttEvans2012} of the sources detectable (in imaging and in spectroscopy) by the proposed deep survey as a function of the redshift. Also shown, for comparison, are the SFRs associated to the minimum luminosities represented in the IR luminosity functions at several redshifts determined by \citet{Gruppioni13} on the basis of \textit{Herschel}/PACS and SPIRE surveys. The improvement over \textit{Herschel} is impressive. The deep survey will sample SFRs well below those of the most efficient star formers, estimated to be $\simeq 100\,M_\odot$/yr \citep{ForsterSchreiber2006,Cai13}. It will therefore allow a full reconstruction of the dust obscured cosmic star formation history up to high redshifts.

%%%%%%%%%%%%

\begin{figure*} % Fig. 6
  \includegraphics[trim=0.8cm 4.7cm 1.3cm 5.2cm,clip=true,width=\textwidth]{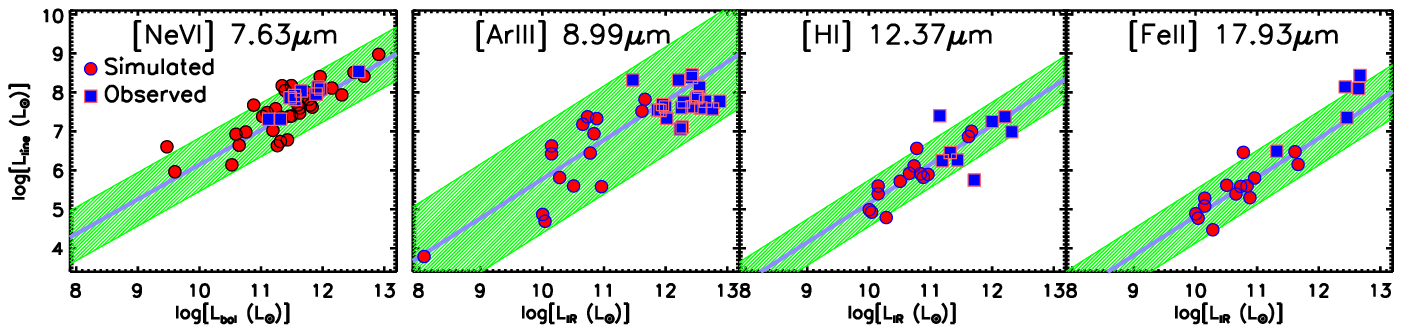} %\vspace{-0.5cm}
  \caption{Comparison between observed (blue squares) and expected (red circles) line luminosities, obtained using ITERA (see
    Section\,\ref{sect:line_vs_IR} for details). The [NeVI]\-7.63\,$\mu$m data (for AGNs) comes from the \citet{Sturm02} and
    \citet{Veilleux09} catalogues. For the other 3 lines the observed line luminosities (for star-forming galaxies) are from
    \citet{Willett11}. The green bands show the $\pm2\,\sigma$ dispersion around the line to continuum luminosity relations derived with ITERA.}
  \label{fig:itera_calibrations}
\end{figure*}

%%%%%%%
% TABLE %
%%%%%%%%

%%%%%%%%
\begin{table}%[h]
  \centering
  \footnotesize
  \begin{tabular}{lcc}
    \hline
    \hline
    % \rule[-3mm]{0mm}{6mm}
    Spectral line & $\displaystyle\big\langle\log\big(\displaystyle{L_{\ell}\over L_{\rm IR}}\big)\big\rangle\ $ & $\sigma$ \\
    \hline
    ${\rm H_{2}}5.51\mu$m		& -4.29	& 0.39	\\
    ${\rm PAH}6.2\mu$m$^{2}$	& -2.20	& 0.36	\\
    ${\rm H_{2}}6.91\mu$m		& -3.97	& 0.39	\\
    ${\rm [ArII]}6.98\mu$m		& -3.96	& 0.32	\\
    ${\rm PAH}7.7\mu$m$^{2}$	& -1.64	& 0.36	\\
    ${\rm PAH}8.6\mu$m$^{2}$	& -2.16	& 0.36	\\
    ${\rm [ArIII]}8.99\mu$m		& -4.22	& 0.69	\\
    ${\rm H_{2}}9.66\mu$m$^{2}$	& -3.96	& 0.52	\\
    ${\rm [SIV]}10.49\mu$m$^{2}$	& -3.95	& 0.69	\\
    ${\rm PAH}11.3\mu$m$^{1}$	& -2.29	& 0.36	\\
    ${\rm H_{2}}12.28\mu$m$^{2}$	& -4.12	& 0.54	\\
    ${\rm HI}12.37\mu$m		& -4.85	& 0.30	\\
    ${\rm [NeII]}12.81\mu$m$^{1}$	& -3.11	& 0.45	\\
    ${\rm [ClII]}14.38\mu$m		& -5.44	& 0.33	\\
    ${\rm [NeIII]}15.55\mu$m$^{1}$	& -3.69	& 0.47	\\
    ${\rm H_{2}}17.03\mu$m$^{1}$	& -4.04	& 0.46	\\
    ${\rm [FeII]}17.93\mu$m		& -5.18	& 0.34	\\
    ${\rm [SIII]}18.71\mu$m$^{1}$	& -3.49	& 0.48	\\
    ${\rm [ArIII]}21.82\mu$m	& -5.40	& 0.70	\\
    ${\rm [FeIII]}22.90\mu$m	& -6.56	& 0.33	\\
    ${\rm [FeII]}25.98\mu$m		& -4.29	& 0.44	\\
    ${\rm [SIII]}33.48\mu$m$^{1}$	& -3.05	& 0.31	\\
    ${\rm [SiII]}34.82\mu$m$^{1}$	& -2.91	& 0.28	\\
    \hline
    \multicolumn{2}{l}{\scriptsize{$^{1}$Taken from \citet{Bonato2014a}}}\\
    \multicolumn{2}{l}{\scriptsize{$^{2}$Taken from \citet{Bonato2014b}}}\\
    \hline
    \hline
  \end{tabular}
  \caption{Mean values of the log of line to IR (8-1000\,$\mu$m) continuum luminosities for star-forming galaxies, $\langle\log({L_{\ell}/L_{\rm IR}})\rangle$, and associated dispersions $\sigma$. For the PAH$\, 11.3\,\mu$m and H$_{2}\, 17.03\,\mu$m lines, the tabulated values, taken from \citet{Bonato2014a}, have been computed excluding local ULIRGs, for which the luminosity in these lines was found to be uncorrelated with $L_{\rm IR}$; the mean values of $\log(L_{\ell})$ in these two lines for local ULIRGs, $\log(L_{\ell}/L_\odot)$, were found to be $9.01$ and $8.07$ (with dispersions of $0.28$ and $0.34$) respectively.}
  \label{tab:sb_c_values}
\end{table}

%%%%%%%
% TABLE %
%%%%%%%%
\begin{table}%[h]
  \centering
  \footnotesize
  \begin{tabular}{lccc}
    \hline
    \hline
    % \rule[-2mm]{0mm}{4mm}
    Spectral line & $a$ & $b$ & disp ($1\sigma$)\\
    \hline
    ${\rm [MgVIII]}3.03\mu$m	& 0.73	& -1.65	& 0.60	\\
    ${\rm [CaIV]}3.21\mu$m		& 0.89	& -2.90	& 0.31	\\	
    ${\rm [SiIX]}3.92\mu$m		& 0.65	& -1.77	& 0.88	\\
    ${\rm [CaV]}4.20\mu$m		& 0.90	& -4.00	& 0.34	\\
    ${\rm [MgIV]}4.49\mu$m		& 0.89	& -3.01	& 0.32	\\
    ${\rm [ArVI]}4.52\mu$m		& 0.83	& -3.36	& 0.34	\\
    ${\rm H_{2}}5.51\mu$m		& 1.23	& -7.01	& 0.40	\\
    ${\rm [MgV]}5.60\mu$m		& 0.91	& -2.94	& 0.34	\\
    ${\rm [SiVII]}6.50\mu$m		& 0.83	& -3.55	& 0.37	\\
    ${\rm H_{2}}6.91\mu$m		& 0.80	& -2.40	& 0.34	\\
    ${\rm [ArII]}6.98\mu$m		& 0.84	& -4.21	& 0.64	\\
    ${\rm [NeVI]}7.63\mu$m		& 0.79	& -1.48	& 0.42	\\
    ${\rm [ArV]}7.90\mu$m		& 0.87	& -3.85	& 0.32	\\
    ${\rm [ArIII]}8.99\mu$m		& 0.98	& -4.15	& 0.37	\\
    ${\rm H_{2}}9.66\mu$m$^{1}$	& 1.07	& -5.32	& 0.34	\\
    ${\rm [SIV]}10.49\mu$m$^{1}$	& 0.90	& -2.96	& 0.24	\\
    ${\rm [CaV]}11.48\mu$m		& 0.90	& -5.12	& 0.34	\\
    ${\rm H_{2}}12.28\mu$m$^{1}$	& 0.94	& -3.88	& 0.24	\\
    ${\rm HI}12.37\mu$m		& 0.86	& -3.84	& 0.34	\\
    ${\rm [NeII]}12.81\mu$m$^{1}$	& 0.98	& -4.06	& 0.37	\\
    ${\rm [ArV]}13.09\mu$m		& 0.87	& -3.85	& 0.32	\\
    ${\rm [MgV]}13.50\mu$m		& 0.91	& -4.01	& 0.34	\\
    ${\rm [NeV]}14.32\mu$m$^{1}$	& 0.78	& -1.61	& 0.39	\\
    ${\rm [ClII]}14.38\mu$m		& 0.85	& -4.83	& 0.57	\\
    ${\rm [NeIII]}15.55\mu$m$^{1}$	& 0.78	& -1.44	& 0.31	\\
    ${\rm H_{2}}17.03\mu$m$^{1}$	& 1.05	& -5.10	& 0.42	\\
    ${\rm [FeII]}17.93\mu$m		& 0.84	& -3.80	& 0.54	\\
    ${\rm [SIII]}18.71\mu$m$^{1}$	& 0.96	& -3.75	& 0.31	\\
    ${\rm [ArIII]}21.82\mu$m	& 0.98	& -5.34	& 0.36	\\
    ${\rm [FeIII]}22.90\mu$m	& 0.79	& -4.85	& 0.60	\\
    ${\rm [NeV]}24.31\mu$m$^{1}$	& 0.69	& -0.50	& 0.39	\\
    ${\rm [OIV]}25.89\mu$m$^{1}$	& 0.70	& -0.04	& 0.42	\\
    ${\rm [FeII]}25.98\mu$m		& 0.87	& -3.71	& 0.55	\\
    ${\rm [SIII]}33.48\mu$m$^{1}$	& 0.62	& 0.35	& 0.30	\\
    ${\rm [SiII]}34.82\mu$m		& 0.89	& -3.14	& 0.52	\\
    \hline
    \multicolumn{2}{l}{\scriptsize{$^{1}$Taken from \citet{Bonato2014b}}}\\
    \hline
    \hline
  \end{tabular}
  \caption{Coefficients of the best-fit linear relations between line and AGN bolometric luminosities, $\log({L_{\ell}})=a\cdot\log({L_{\rm bol}})+b$, and $1\sigma$ dispersions associated to the relations.}
  \label{tab:agn_a_b_values}
\end{table}

\begin{figure*} % Fig. 7
  \includegraphics[trim=1.25cm 1.4cm 2.1cm 2.2cm,clip=true,width=\textwidth]{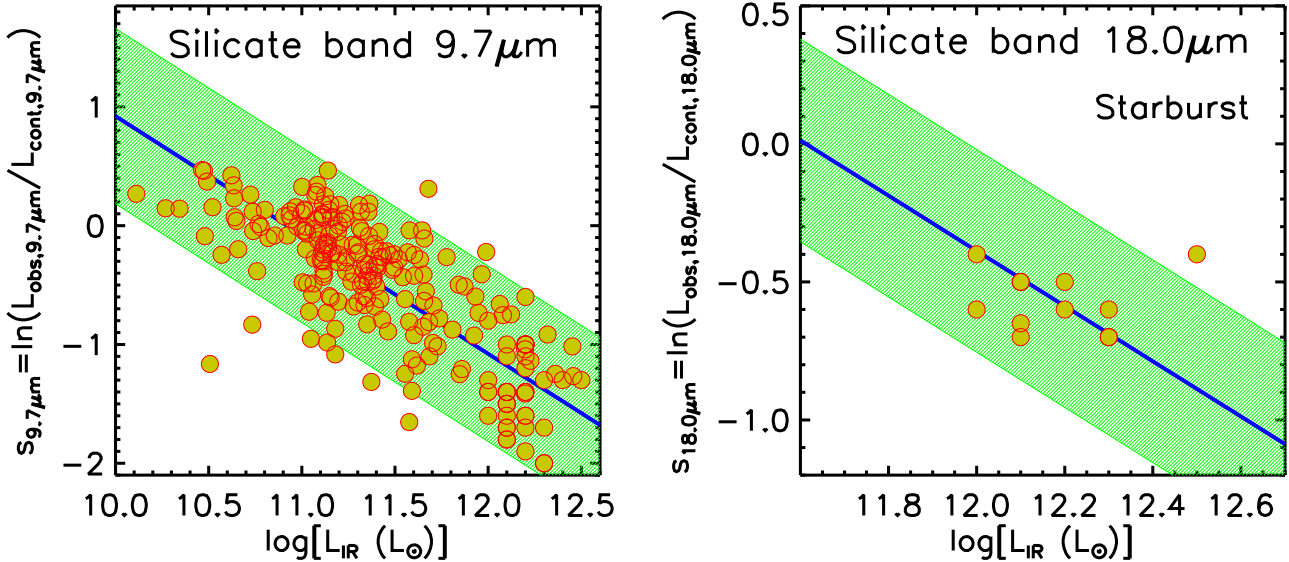} %\vspace{-0.5cm}
  \caption{Strength of the 9.7 and 18.0$\,\mu$m silicate bands, as a function of the IR luminosity, for star-forming galaxies. Data from \citet{Stierwalt13}, \citet{Imanishi07}, \citet{Imanishi09}, and \citet{Imanishi10}. The green bands show the $2\,\sigma$ intervals around the mean linear relations $s_{\lambda} = -\log(L_{\rm IR}) + c$ (blue lines).}
  \label{fig:cal_silicate_sb}
\end{figure*}

\begin{figure*} % Fig. 8
  \includegraphics[trim=0.0cm 3.0cm 1.0cm 3.8cm,clip=true,width=\textwidth]{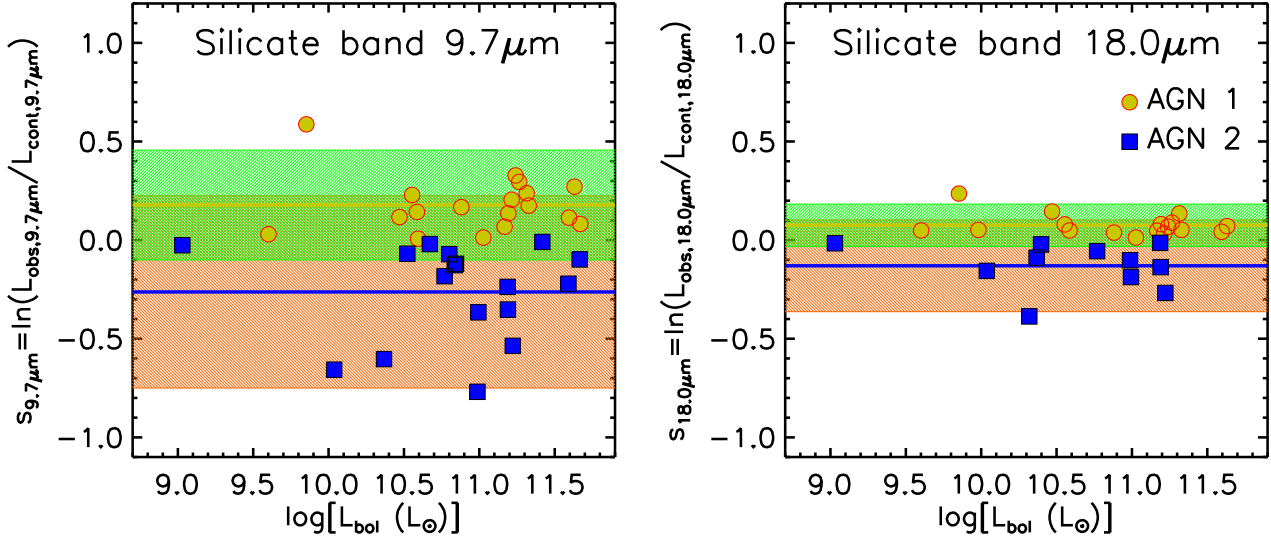} %\vspace{-0.5cm}
  \caption{Strength of the 9.7 and 18.0$\,\mu$m silicate bands as a function of the bolometric luminosity for type\,1 (circles) and type\,2 (squares) AGNs from the \citet{Gallimore10} sample. The green or orange bands show the $2\,\sigma$ spreads around the mean strength for type\,1's (yellow lines) or type\,2's (blue lines), respectively. }
  \label{fig:cal_silicate_agn}
\end{figure*}

%%%%%%%%%%%%
% FIGURE %
%%%%%%%%%%%%
\begin{figure*} % Fig. 9
  \makebox[\textwidth][c]{
    \includegraphics[trim=0.2cm 1.9cm 1.1cm
    2.6cm,clip=true,width=\textwidth, angle=0]{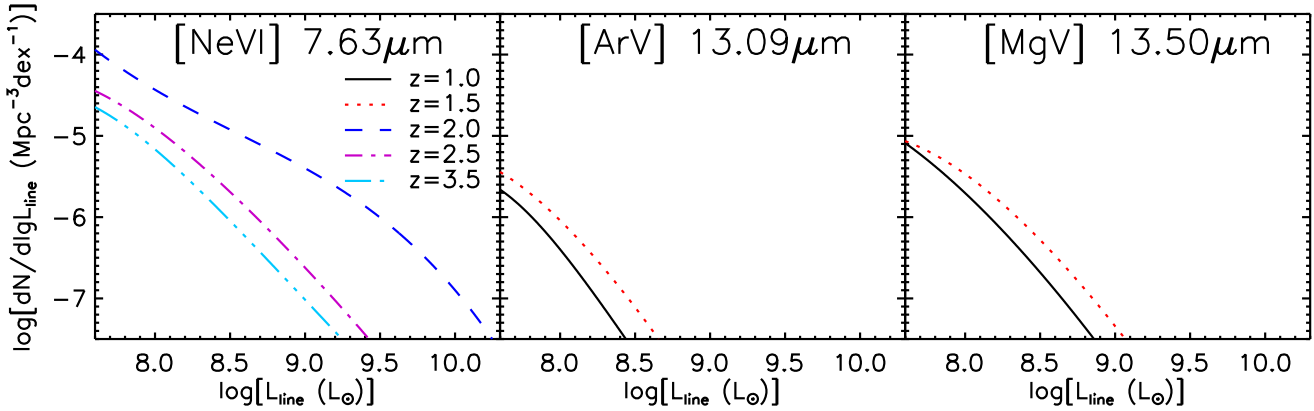}
  }
  \caption{Predicted luminosity functions of the AGN lines [NeVI]\,$7.63$, [ArV]\,$13.09$, and [MgV]\,$13.50\,\mu$m at
    different redshifts at which these lines can be detected by the SMI spectrometers. }
  \label{fig:LF_agn_MIS}
\end{figure*}
%

%%%%%%%%%%%%
% FIGURE %
%%%%%%%%%%%%
\begin{figure*} % Fig. 10
  \makebox[\textwidth][c]{
    \includegraphics[trim=-0.2cm 2.2cm 0.6cm
    2.5cm,clip=true,width=\textwidth, angle=0]{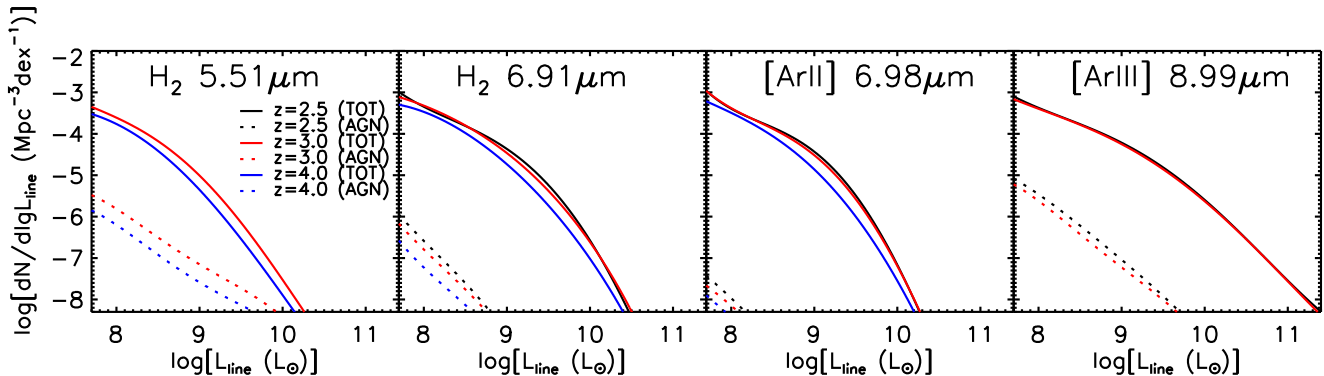}
  }
  \caption{Examples of predicted luminosity functions for the total (starburst plus AGN) emission (solid lines) and for the AGN
    component only (dotted lines) of some lines of our sample, at different redshifts at which these lines can be detected by the
    SMI spectrometers. }
  \label{fig:LF_tot_MIS}
\end{figure*}
%

%%%%%%%%%%%%
% FIGURE %
%%%%%%%%%%%%
\begin{figure*} % Fig. 11
  \hspace{+0.0cm} \makebox[\textwidth][c]{
    % \includegraphics[width=1.55\textwidth,
    % angle=0]{../LFIR/Grafici_articolo/panel_intcounts.ps}
    \includegraphics[trim=2.8cm -0.4cm 3.1cm
    0.0cm,clip=true,width=1.0\textwidth,
    angle=0]{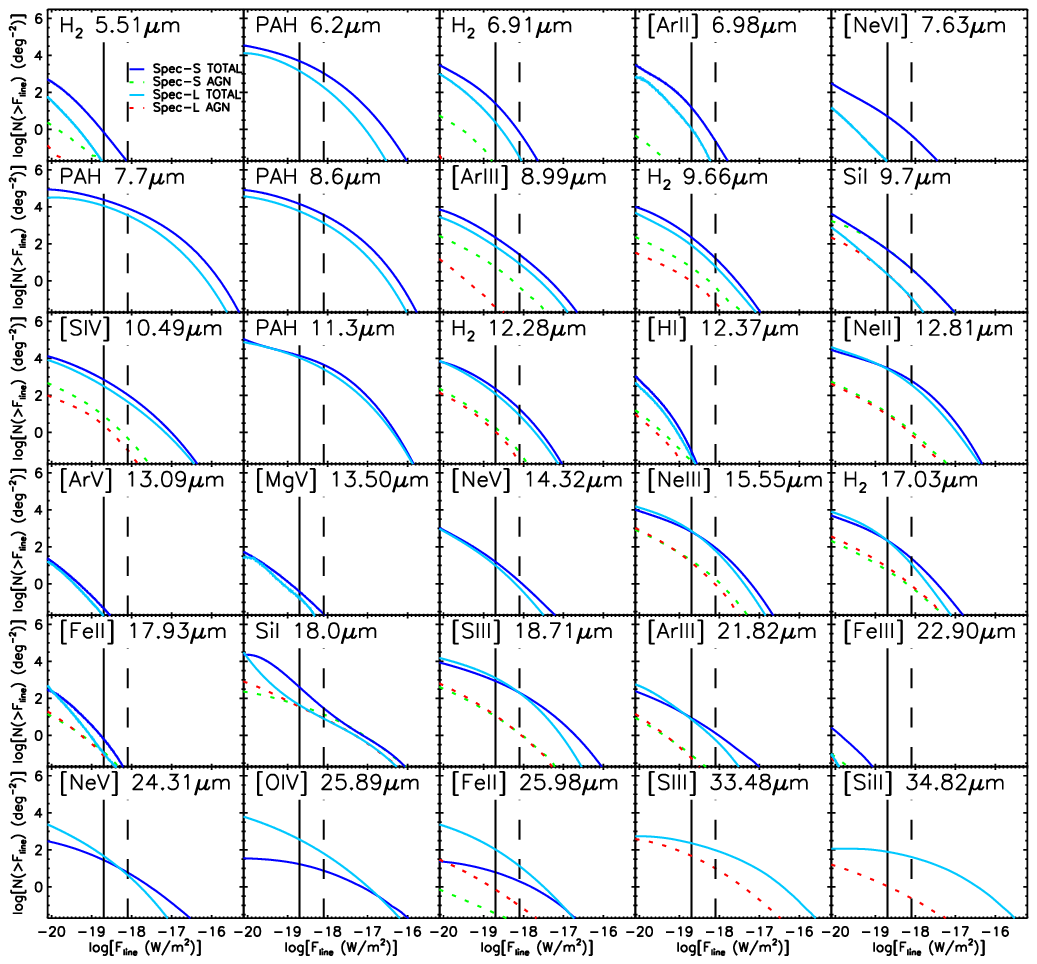}
  }
  % \vspace{-4.5cm}
  \caption{Integral counts in 30 MIR lines of galaxies as a whole (starburst plus AGN components; solid lines) and of AGNs only (dotted lines) over the Spec-S and the Spec-L wavelength ranges. The vertical lines correspond to the detection limits ($2.0\cdot10^{-19}\,\hbox{W}/\hbox{m}^{2}$ for Spec-S and $8.0\cdot10^{-19}\,\hbox{W}/\hbox{m}^{2}$ for Spec-L) for 1-h  exposure per FoV.}
  \label{fig:intcounts_spica_all_spec}
\end{figure*}

%%%%%%%%%%%%%%%%%%%%%%%%%%%%%%%%%%%%%%%%%%%%%%%%%%%%%%%%%%%%%%%%%%%%%%%%%%%%%
\section{Line versus IR luminosity}\label{sect:line_vs_IR}
%%%%%%%%%%%%%%%%%%%%%%%%%%%%%%%%%%%%%%%%%%%%%%%%%%%%%%%%%%%%%%%%%%%%%%%%%%%%%

To estimate the counts of galaxy and AGN line detections by SMI surveys we coupled the redshift dependent IR (in the case of galaxies) or bolometric (in the case of AGNs) luminosity functions of the source populations with relationships between line and IR or bolometric luminosities. We have considered the following set of 41 IR lines:
\begin{itemize}
\item 3 coronal region lines: [MgVIII]\-3.03, [SiIX]\-3.92 and [SiVII]\-6.50$\,\mu$m;
\item 13 AGN fine-structure emission lines: [CaIV]\-3.21, [CaV]\-4.20, [MgIV]\-4.49, [ArVI]\-4.52, [MgV]\-5.60, [NeVI]\-7.63, [ArV]\-7.90, [CaV]\-11.48, [ArV]\-13.09, [MgV]\-13.50, [NeV]\-14.32, [NeV]\-24.31 and [OIV]\-25.89$\,\mu$m;
\item 19 fine-structure emission lines that can also be produced in star-formation regions:
\begin{itemize}
\item 10 stellar/HII region lines: [ArII]\-6.98, [ArIII]\-8.99, [SIV]\-10.49, HI\-12.37, [NeII]\-12.81, [ClII]\-14.38, [NeIII]\-15.55, [SIII]\-18.71, [ArIII]\-21.82 and [SIII]\-33.48\,$\mu$m;
\item 4 lines from photodissociation regions: [FeII]\-17.93, [FeIII]\-22.90, [FeII]\-25.98 and [SiII]\-34.82$\,\mu$m;
\item 5 molecular hydrogen lines: H$_{2}$\,\-5.51, H$_{2}$\,\-6.91, H$_{2}$\,\-9.66, H$_{2}$\,\-12.28 and H$_{2}$\,\-17.03\,$\mu$m;
\end{itemize}
\item 4 Polycyclic Aromatic Hydrocarbon (PAH) lines at $6.2$, $7.7$, $8.6$ and $11.3\,\mu$m;
\item the 2 emission and absorption silicate bands at $9.7$ and $18.0\,\mu$m.
\end{itemize}
For the PAH\,\-6.2, PAH\,\-7.7, PAH\,\-8.6, PAH\,\-11.3, H$_{2}$\,\-9.66, [SIV]\-10.49, H$_{2}$\,\-12.28, [NeII]\-12.81, [NeV]\-14.32, [NeIII]\-15.55, H$_{2}$\,\-17.03, [SIII]\-18.71, [NeV]\-24.31, [OIV]\-25.89, [SIII]\-33.48 and [SiII]\-34.82\,$\mu$m lines we have used the relationships derived by \citet{Bonato2014a,Bonato2014b} on the basis of observations collected from the literature. %The results are summarized in Table\,\ref{tab:sb_c_values} for star forming lines and in Table\,\ref{tab:agn_a_b_values} for AGN lines.

For all the other lines, with either missing or insufficient data, the line to continuum luminosity relations were derived using the IDL Tool for Emission-line Ratio Analysis (ITERA)\footnote{http://home.strw.leidenuniv.nl/~brent/itera.html} written by Brent Groves. ITERA uses the library of published photoionization and shock models for line emission of astrophysical plasmas produced by the Modelling And Prediction in PhotoIonised Nebulae and Gasdynamical Shocks (MAPPINGS III) code.

Among the options offered by ITERA we have chosen, for starbursts, the \citet{Dopita06} models and, for AGNs, the dust free isochoric narrow line region (NLR) models for type\,1's and the dusty radiation-pressure dominated NLR models for type\,2's \citep{Groves04}. The chosen models are those which provide the best overall fit (minimum $\chi^2$) to the observed line ratios of local starbursts in the \citet{Bern09} catalogue and of AGNs in the sample built by \citet{Bonato2014b} combining sources from the \citet{Tomm08,Tomm10}, \citet{Sturm02}, and \citet{Veilleux09} catalogues.

As an example we compare, in Fig.~\ref{fig:itera_calibrations}, the line luminosities as a function of IR luminosities obtained using ITERA with the observed ones (the majority of which were published after the \citealt{Groves04} and \citealt{Dopita06} models) for  [NeVI]\-7.63, [ArIII]\-8.99, HI\-12.37 and [FeII]\-17.93\,$\mu$m. As a further test, we have compared the luminosities of the fainter high-ionization lines measured by \citet{Spinoglio2005} in the prototype Seyfert~2 galaxy NGC 1068 with those obtained via ITERA. Adopting for the active nucleus of this object a bolometric luminosity of $3\times 10^{11}\,L_\odot$ \citep{Bock2000} we find, from ITERA, $\log(L_\ell)$ in the ranges  ($2\,\sigma$) [6.13,7.33],[4.81,6.57],[6.88,7.52],[5.83,6.51],[7.16,7.84] for [MgVIII]3.03, [SiIX]3.92, [MgIV]4.49, [ArVI]4.52 and [MgV]5.60\,$\mu$m, respectively, in reasonably good agreement with the corresponding measured values (6.95, 6.60, 6.79, 7.08 and 7.16, respectively), especially taking into account the substantial uncertainty in the estimated bolometric luminosity. There is no indication of systematic over- or under-estimate of the line luminosities.

Note that our counts take into account the absorption of the fine-structure lines near the strong silicate absorption features, since such absorption is properly dealt with by ITERA.

The data on starburst galaxies are consistent with a direct proportionality between line and IR luminosity. The mean line to IR luminosity ratios, $\langle \log(L_{\ell}/L_{\rm IR})\rangle$, and the dispersions, $\sigma$, around them are listed in Table\,\ref{tab:sb_c_values}. In the case of AGNs the data are described by linear mean relations  $\log({L_{\ell}})=a\cdot\log({L_{\rm bol}})+b$. The coefficients of such relations and the dispersions around them are listed in Table~\ref{tab:agn_a_b_values}.

From a practical viewpoint the faintest lines considered above will hardly be detected by SPICA instruments. Nevertheless we thought it useful to derive relationships with the continuum luminosity for as many IR lines as possible to provide a tool to estimate exposure times to detect such lines with pointed observations, not necessarily only with SPICA instruments.

For the 9.7 and 18.0$\,\mu$m silicate bands we have used the observed correlations between the IR luminosity and the relative
strength of the features, defined \citep[see, e.g.,][]{Spoon07} as the natural logarithm of the ratio between the observed flux
density at the center of the silicate feature, $F_{{\rm obs}, \lambda_{f}(1+z)}$, and the local continuum flux density, $F_{{\rm cont}, \lambda_{f}(1+z)}$,
\begin{eqnarray}
  s_{\lambda_{f}} = \ln{\frac{F_{{\rm obs}, \lambda_{f}(1+z)}}{F_{cont, \lambda_{f}(1+z)}}} = \ln{\frac{L_{{\rm obs}, \lambda_{f}}}{L_{{\rm cont}, \lambda_{f}}}}
\end{eqnarray}
where $\lambda_{f}$ is the rest-frame wavelength of the feature (i.e. 9.7 or 18\,$\mu$m), while $L_{{\rm obs}, \lambda_{f}}$ and $L_{{\rm cont}, \lambda_{f}}$ are the corresponding (monochromatic) luminosities at that wavelength.

To calibrate the relationships between the silicate band strength and the IR luminosity for the starburst component we have used data from \citet[][only for 9.7$\mu$m silicate band]{Stierwalt13}, excluding the objects with low $6.2\,\mu$m PAH equivalent widths ($\hbox{EQW}_{6.2\mu\rm m}<0.27\mu$m) whose MIR emission is likely to be substantially contaminated by an AGN, and the starburst dominated galaxies from \citet{Imanishi07}, \citet{Imanishi09}, and \citet{Imanishi10} catalogues (these authors actually provide optical depths, $\tau_{\lambda}=-s_{\lambda}$).

As illustrated by Fig.\,\ref{fig:cal_silicate_sb}, the 9.7\,$\mu$m strength of starburst galaxies shows a clear linear anti-correlation with the log of the IR (8-1000$\mu$m) luminosity, consistent with a constant $c=s_{\lambda}+\log{L_{IR}}$ with a mean value $\langle c\rangle=10.92$ and dispersion of $0.37$. The linear correlation coefficient is $-0.78$, corresponding to a correlation significant at the $15.7\,\sigma$ level.

The very few data on the 18.0\,$\mu$m strength do not show any significant correlation with $L_{\rm IR}$. However, given the poor statistics, the possibility of a correlation cannot be ruled out either. If, in analogy to the 9.7\,$\mu$m strength, we assume also for the 18.0\,$\mu$m one a relation of the form $s_{\lambda}=-\log{L_{IR}}+c$ we get $\langle c\rangle=11.61$ with a dispersion of $0.18$. If, instead, the two quantities are uncorrelated the data give  $\langle s_{\lambda}\rangle=-0.57$ with a dispersion of 0.11. In the latter case, the number of galaxies detectable in that band by the SPICA SMI spectrometers in 1\,hour integration per FoV decreases by a factor $\sim6$.

For AGNs we have used the \citet{Gallimore10} sample, neglecting the silicate absorption for type\,1's and the emission for type\,2's. As illustrated by Fig.~\ref{fig:cal_silicate_agn} the silicate strengths of AGNs appear to be uncorrelated with the bolometric luminosities. We have therefore adopted Gaussian distributions of $s_{\lambda}$ around mean values, $\langle s_{\lambda}\rangle$, independent of $L_{\rm bol}$. We have obtained: $\langle s_{9.7\mu\rm m}\rangle_{\rm AGN\,1}=0.18$, $\langle s_{9.7\mu\rm m}\rangle_{\rm AGN\,2}=-0.26$, $\langle s_{18.0\mu\rm m}\rangle_{\rm AGN\,1}=0.08$ and $\langle s_{18.0\mu\rm m}\rangle_{\rm AGN\,2}=-0.13$ with dispersions of 0.14, 0.24, 0.05, and 0.12, respectively.

The line luminosity functions have been computed starting from the redshift-dependent IR luminosity functions given by the evolutionary model, including both the starburst and the AGN component. To properly take into account the dispersion in the relationships between line and continuum luminosities we have used the Monte Carlo approach described in \citet{Bonato2014b}. Examples of line luminosity functions at various redshifts are shown in Figs.~\ref{fig:LF_agn_MIS} and \ref{fig:LF_tot_MIS}.

%%%%%%%%%%%%
% FIGURE %
%%%%%%%%%%%%
\begin{figure} % Fig. 12
  %\makebox[\textwidth][c]{
    \includegraphics[trim=0.0cm 0.0cm 0.7cm 0.7cm,clip=true,width=0.5\textwidth,
    angle=0]{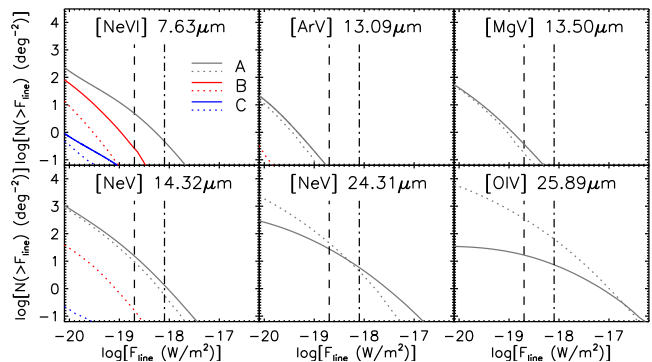}
%  }
  \caption{Contributions of different AGN populations to the SMI Spec-S (solid lines) and Spec-L (dotted lines) integral counts in 6 AGN lines ([NeVI]\-7.63, [ArV]\-13.09, [MgV]\-13.50, [NeV]\-14.32, [NeV]\-24.31 and [OIV]\-25.89$\,\mu$m). A: AGNs associated to late-type galaxies plus optically selected AGNs; B: AGNs associated to unlensed proto-spheroids; C: AGNs associated to strongly gravitationally lensed proto-spheroids. The vertical lines correspond to the detection limits ($2.0\cdot10^{-19}\,\hbox{W}/\hbox{m}^{2}$ for Spec-S and $8.0\cdot10^{-19}\,\hbox{W}/\hbox{m}^{2}$ for Spec-L) for 1-h  exposure per FoV.}
  \label{fig:intcounts_spica}
\end{figure}

%%%%%%%%%%%%
% FIGURE %
%%%%%%%%%%%%
\begin{figure} % Fig. 13
  \hspace{+0.0cm} %\makebox[\textwidth][c]{
    % \includegraphics[width=1.55\textwidth,
    % angle=0]{../LFIR/Grafici_articolo/panel_redshdistr.ps}
    \includegraphics[trim=0.2cm 0.3cm 0.8cm 1.0cm,clip=true,width=0.5\textwidth,
    angle=0]{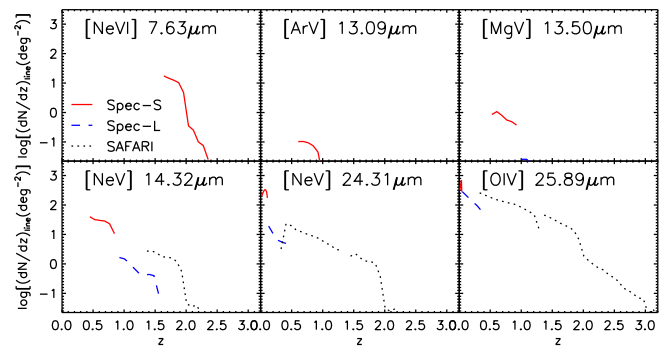}
 % }
  % \vspace{-3.cm}
  \caption{Predicted redshift distributions of AGNs detected in 6 lines ([NeVI]\-7.63, [ArV]\-13.09, [MgV]\-13.50, [NeV]\-14.32,  [NeV]\-24.31 and [OIV]\-25.89$\,\mu$m) detected by SMI Spec-S (solid red lines) and Spec-L (dashed blue lines), compared, for three of them ([NeV]\-14.32, [NeV]\-24.31 and [OIV]\-25.89$\,\mu$m), with those obtained in \citet{Bonato2014b} for SPICA/SAFARI (dotted black lines), for an 1\,h exposure per FoV.}
  \label{fig:redshiftdistr_spica}
\end{figure}

%The [MgVIII]\-3.03, [CaIV]\-3.21, [SiIX]\-3.92, [CaV]\-4.20, [MgIV]\-4.49, [ArVI]\-4.52, [MgV]\-5.60, [SiVII]\-6.50, [ArV]\-7.90, [CaV]\-11.48 and [ClII]\-14.38$\mu$m lines, listed in Tables \,\ref{tab:sb_c_values} and \,\ref{tab:agn_a_b_values}, are not considered in the following analysis because they are undetectable by SMI ({\bf Mattia: se e' cosi' perche' abbiamo considerato queste righe nelle tabelle? Forse e' il caso di rimuoverle?}).

%%%%%%%%%%%%%%%%%%%%%%%%%%%%%%%%%%%
\section{Line luminosity functions and number counts}\label{sect:LF}
%%%%%%%%%%%%%%%%%%%%%%%%%%%%%%%%%%%

%
Our predictions for the integral counts in both the Spec-S and the Spec-L channels, for 30 lines are shown in Fig.~\ref{fig:intcounts_spica_all_spec}. %The predicted redshift distributions of sources detected per square degree in a number of lines by the Spec-S and the Spec-L for exposures of 1\,h per FoV are presented in Tables \ref{tab:counts_spica1a_s} and \ref{tab:counts_spica1a_l}, respectively.
In 1\,h integration per FoV, the Spec-S will detect $\sim 70$ AGN lines per square degree, primarily the [NeV]$24.31\,\mu$m line ($\sim 40\%$ of detections); the Spec-L will detect mostly the [OIV]$25.89\,\mu$m line (over 90\% of the $\sim 70$ AGN line detections). The $\simeq 140$ lines are produced by $\simeq 110$ individual AGNs. Only a handful of them are optically selected. Thus this survey will be very efficient at selecting obscured AGNs, complementing optical surveys.

Figure~\ref{fig:intcounts_spica} shows the contributions of AGNs associated with different galaxy populations to the SMI Spec-S and Spec-L counts in six AGN lines. The redshift distributions of AGNs detected in each of these lines in 1\,h integration/FoV are displayed in Fig.~\ref{fig:redshiftdistr_spica}. For three of the lines ([NeV]\-14.32, [NeV]\-24.31 and [OIV]\-25.89\,$\mu$m) we also show, for comparison, the redshift distributions obtained by \citet{Bonato2014b} for a SPICA/SAFARI survey  (again for a 1\,h exposure per FoV). The two instruments cover nicely complementary redshift intervals.

\begin{figure} % Fig. 14
%  \makebox[\textwidth][c]{
    \includegraphics[trim=2.3cm 0.4cm 0.8cm  0.2cm,clip=true,width=0.5\textwidth, angle=0]{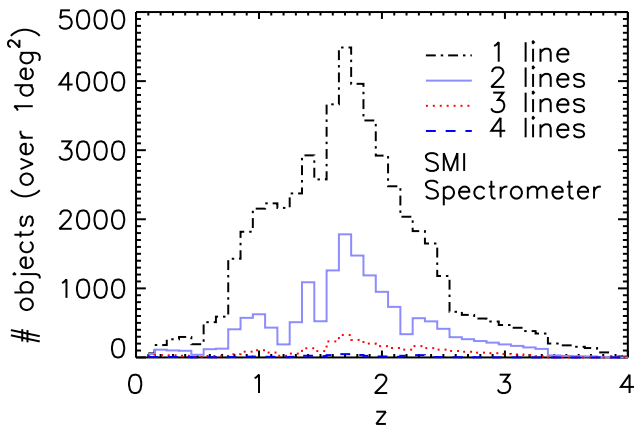}
%    \includegraphics[trim=2.3cm 0.4cm 0.8cm 0.2cm,clip=true,width=0.5\textwidth,  angle=0]{Lir_distr_tot_hybrid_MIS.eps}
%  }
  \caption{Predicted redshift distributions of galaxies (starburst plus AGN components) detectable in one (black histogram), two (cyan), three (red) and four (blue) spectral lines, by a SPICA SMI spectrometer (Spec-S plus Spec-L) survey covering $1\,\hbox{deg}^2$ in 1\,h integration/FoV.}
  \label{fig:multiple_MIS}
\end{figure}

Figure~\ref{fig:multiple_MIS} illustrates the redshift distributions of sources for which the SMI spectrometers will detect at least 1, 2, 3 or 4 lines in 1\,h integration per FoV over an area of $1\,\hbox{deg}^2$. The corresponding numbers of detected lines are  about 53,000, 16,000, 3,100, 360, 55 and 8, respectively. Sources detected in at least one line include $\sim 200$ strongly lensed galaxies at $z>1$. 

Note that the $5\,\sigma$ detection of only one line does not necessarily imply that the redshift determination is problematic. Many of the objects detected in only one line can show a suite of weak but measurable features (other lines, absorptions, PAH bumps). The global \textit{pattern} can then allow the determination of reliable redshifts even when individual features are only significant at the 2--$3\,\sigma$ level. 

\begin{figure}  % Fig. 15
  \hspace{+0.0cm} %\makebox[\textwidth][c]{
    \includegraphics[trim=2.5cm 0.5cm 0.6cm
    0.3cm,clip=true,width=0.5\textwidth, angle=0]{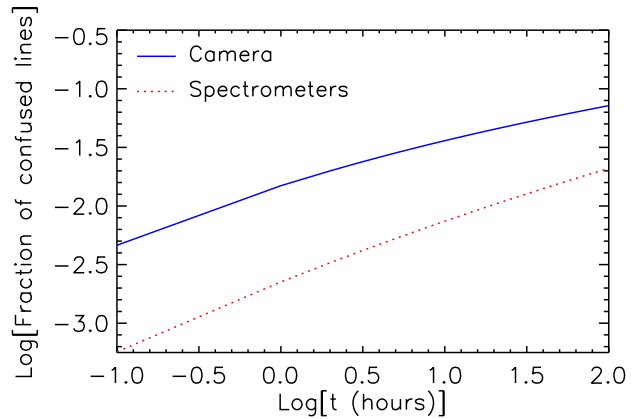}
%  }
  % \vspace{-4.5cm}
  \caption{Detection of multiple spectral lines produced by galaxies at different redshifts falling within the same resolution element: the fraction of cases of confused lines is shown as a function of the integration time per FoV, for a survey with the SMI camera (solid blue line) and with the SMI spectrometers (dotted red line).}
  \label{fig:confusion_lines}
\end{figure}

Some multiple line detections can be due to different galaxies seen by chance within the same resolution element. We have estimated this confusion effect adopting a typical FWHM of $2''$ and assuming a random galaxy distribution (negligible clustering effects). The fractions of two line detections due to confusion by the SMI camera and by the SMI spectrometers are shown as a function of the integration time per FoV in Fig.~\ref{fig:confusion_lines}. The confused fraction is always small (in particular in the case of SMI spectrometers). For example, with an integration time of 1\,hr two line detections due to confusion are $\simeq 1.5\%$ (for the camera survey) and $\simeq 0.2\%$ (for the spectrometers). The number of confusion cases grows almost linearly with the integration time. Therefore the confused fraction grows as we go to fainter fluxes, where the number counts are flatter, but it is still only $\simeq 3.6\%$ (camera) and $\simeq 0.7\%$ (spectrometers) for an integration time of 10\,h per FoV.

\begin{figure} % Fig. 16
  \hspace{+0.0cm} %\makebox[\textwidth][c]{
    \includegraphics[trim=0.0cm 0.0cm 5.6cm
    3.5cm,clip=true,width=0.5\textwidth, angle=0]{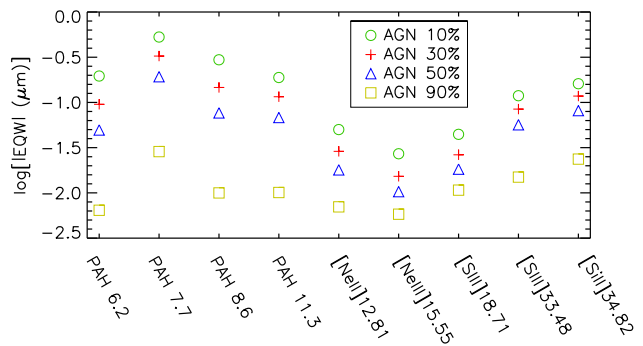}
 % }
  % \vspace{-4.5cm}
  \caption{Equivalent widths (EQWs) of the brightest spectral lines in our sample for a galaxy with a fixed total (star forming plus AGN) IR luminosity of $10^{13} L_{\sun}$ and varying fractions of AGN IR luminosity.}
  \label{fig:EQWs}
\end{figure}

The difference between the AGN and the galaxy SEDs in the SMI range implies that the equivalent widths (EQWs) of the brightest spectral lines excited by star formation are useful indicators of the AGN contribution. This is illustrated by Fig.~\ref{fig:EQWs} which shows the variation of the EQWs of the most prominent star formation lines ($\log(\hbox{EQW})> -1.5$) with the fractional AGN contribution to the total (starburst plus AGN) $L_{IR}$(8-1000$\mu$m). % for a total (starburst plus AGN) IR luminosity of $10^{13}L_{\sun}$.
The PAH lines are particularly effective for this purpose.
\section{Observing strategy}\label{sect:survey}
%%%%%%%%%%%%%%%%%%

As illustrated by Fig.~\ref{fig:intcounts_spica_all_spec}, the integral counts for both SMI spectrometers have a slope flatter than 2 at and below the detection limit for 1\,h integration/FoV for the majority of the lines. Counts of PAH lines with the SMI camera show a similar behaviour (Fig.~\ref{fig:intcounts_spica_all_WFC}).  This means that the number of detections for a fixed observing time generally increases more by extending the survey area than by going deeper, similarly to what found by \cite{Bonato2014a,Bonato2014b} for blind spectroscopic surveys with SPICA/SAFARI.

\begin{figure} % Fig. 17
  % \hspace{+0.0cm}
  %\makebox[\textwidth][c]{
    % \includegraphics[width=1.55\textwidth,
    % angle=0]{../LFIR/Grafici_articolo/panel_intcounts.ps}
    % \includegraphics[trim=0.5cm 2.8cm 1.0cm
    % 3.3cm,clip=true,width=1.1\textwidth,
    % angle=0]{panel_intcounts_all_ccat.eps} }
    % \includegraphics[trim=0.5cm 2.8cm 1.0cm
    % 3.3cm,clip=true,width=1.1\textwidth,
    % angle=0]{panel_intcounts_all_ccat.eps} }
    \includegraphics[trim=2.7cm 0.6cm 3.5cm 0.8cm,clip=true,width=0.23\textwidth,
    angle=0]{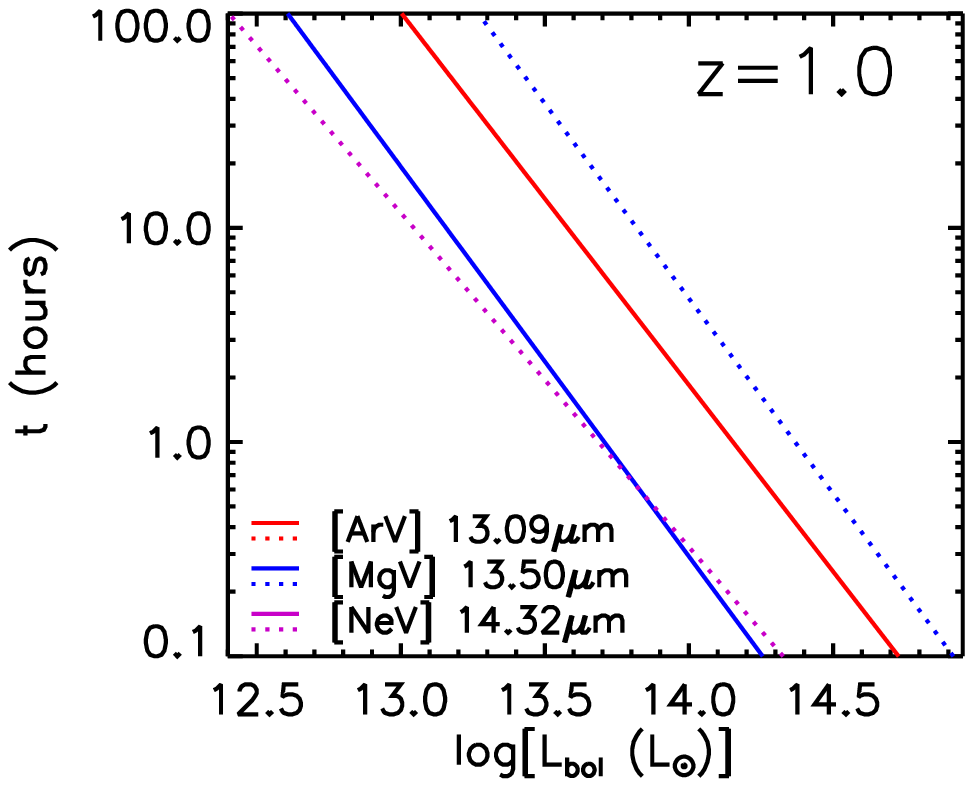}
    \includegraphics[trim=2.7cm 0.6cm 3.5cm  0.8cm,clip=true,width=0.23\textwidth, angle=0]{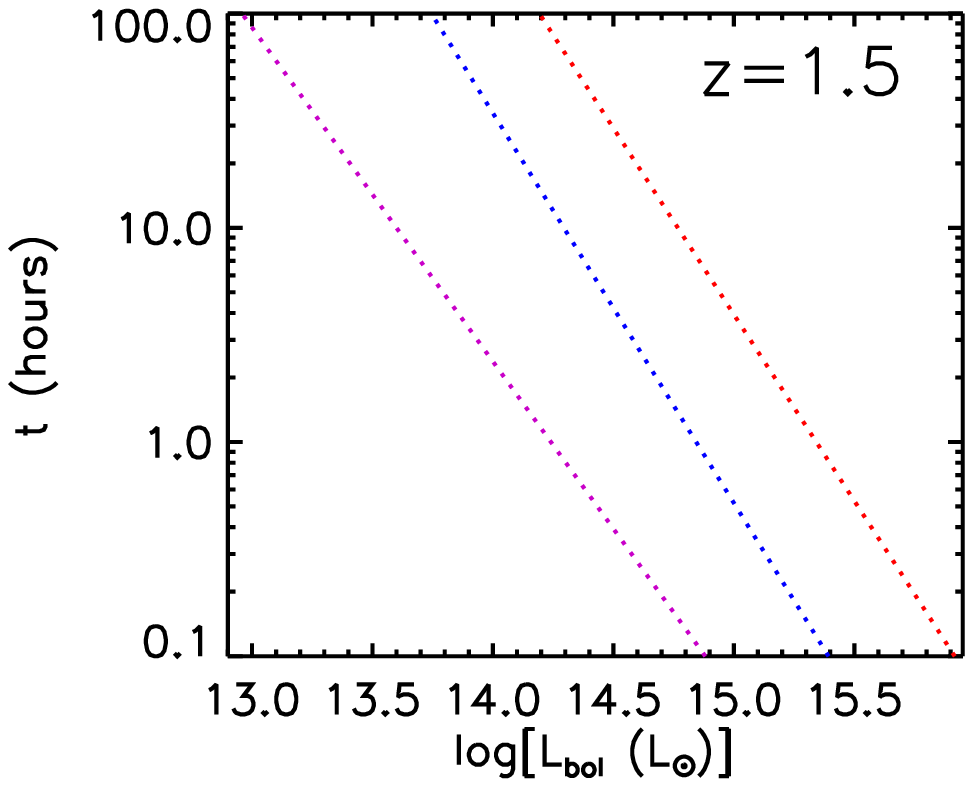}
 % }
  % \vspace{-4.5cm}
  \caption{SPICA SMI Spec-S (solid lines) and Spec-L (dotted lines) exposure time per FoV required for a $5\sigma$ line detection of 3 typical AGN lines ([ArV]\-13.09, [MgV]\-13.50 and [NeV]\-14.32$\mu$m) as a function of the AGN bolometric luminosity  for $z=1.0$ (left) and $z=1.5$ (right).}
  \label{fig:strategy}
\end{figure}

As mentioned in Sect.~\ref{sect:LF}, we expect that a survey of $1\,\hbox{deg}^2$ with 1\,h integration/FoV will detect $\simeq 110$ AGNs. Therefore, to investigate the AGN evolution with sufficient statistics we need a much wider-area.
%
%\textbf{As shown by Fig.~\ref{fig:intcounts_spica_all_WFC}, also the integral counts for the SMI wide field camera have a slope flatter than 2. The proposed survey for the camera is discussed in Section ~\ref{sect:imaging}.}
%
Also, the blind SMI spectroscopic survey may be usefully complemented by follow-up observations of bright high-$z$ galaxies already discovered at (sub-)mm wavelengths over much larger areas. Figure~\ref{fig:strategy} shows the SMI spectrometer exposure time per FoV needed to achieve a $5\sigma$ detection of typical AGN lines at $z=1$ and $z=1.5$ as a function of the bolometric luminosity. We can see, for example, that the [NeV]\-14.32$\,\mu$m line can be detected in 10\,h from an AGN with (real or apparent, i.e. boosted by strong gravitational lensing) bolometric luminosity of $\simeq 10^{13}\,L_\odot$ at $z=1$ and $\simeq 4\times 10^{13}\,L_\odot$ at $z=1.5$. When the South Pole Telescope (SPT) and the \textit{Herschel} survey data will be fully available we will have samples of many hundreds of galaxies with either intrinsic or apparent IR luminosities larger than $10^{13}\,L_{\odot}$ \citep{Neg10,Neg14,Vieira13}. As explained in \citet{Bonato2014b}, pointed observations of those sources can allow us to investigate early phases of the galaxy/AGN co-evolution.

%

%%%%%%%%%%%%
% FIGURE %
%%%%%%%%%%%%

%%%%%%%%%%%%
% FIGURE %
%%%%%%%%%%%%

\section{Conclusions}\label{sect:concl}
%%%%%%%%%%%%%%%%%%

We have worked out predictions for surveys with the SMI wide field camera and with the spectrometers.

The combination of a shallow and of a deep survey with the camera, requiring a total observing time of $\simeq 1046\,$h, will allow an accurate definition of MIR source counts of both galaxies and AGNs over about five decades in flux density, down to $\simeq 3.0\,\mu$Jy, i.e. more than one order of magnitude fainter than the deepest \textit{Spitzer} surveys at $24\,\mu$m. This amounts to resolving almost entirely the MIR extragalactic background. The spectral resolution of the camera is optimally suited to detect PAH lines, yielding redshift measurements. The redshift information will allow us to derive the SFR function down to SFRs hundreds of times lower than was possible using \textit{Herschel} surveys and well below the SFRs of typical star-forming galaxies. The cosmic dust obscured star formation history will then be accurately determined at least up to $z\simeq 4$.

%With an exposure time of 1\,h/FoV, the camera will detect in at least 1 PAH line about 88,000 galaxies per square degree, about 390 of which will be strongly lensed. About 35,000 galaxies will be detected in at least 2 PAH lines.

On the spectroscopic side we have considered 41 MIR lines, 6 of which are predominantly excited by AGN activity while the others are primarily associated with star formation. Relationships between the line luminosity and the IR (for the starburst component) and/or the bolometric luminosity (for the AGN component) are presented. Several of them were derived in previous papers, but many are new.

Using these relationships we computed the expected number counts for the 30 brightest lines. We found that the SMI spectrometers will detect, with an integration time of 1\,h/FoV,  about 52,000 galaxies per square degree in at least one line and about 16,000 in at least two lines. About 200 of galaxies detected in at least one line will be strongly lensed.

%A s 24 MIR fine-structure emission lines partially or entirely excited by AGN activity, as well as in the $6.2$, $7.7$, $8.6$ and $11.3\mu$m PAH lines and in the two silicate bands at $9.7$ and $18.0\mu$m.  The expected outcome of a survey covering $1\,\hbox{deg}^2$ with a 1\,hr integration/FoV is specifically discussed. Moreover we have estimated the number of objects detectable in at least 2, 3, 4, 5 and 6 lines.

The number of expected AGN detections is far lower. For the same integration time (1\,h/FoV) we expect to detect, per square degree, $\simeq 140$ AGN lines, from $\simeq 110$ AGNs. Thus a larger area is necessary to investigate the AGN evolution with good statistics.

Given the low surface density of AGNs detectable by the SMI spectrometers, an efficient way to investigate early phases of the galaxy/AGN co-evolution are pointed observations of the brightest galaxies detected by large area surveys such as those by
\textit{Herschel} and by the SPT.

%%%%%%%%%%%%
% FIGURE %
%%%%%%%%%%%%

\section*{Acknowledgements} We acknowledge financial support from ASI/INAF Agreement 2014-024-R.0 for the {\it Planck} LFI activity of Phase E2 and from PRIN INAF 2012, project ``Looking into the dust-obscured phase of galaxy formation through cosmic zoom lenses in the Herschel Astrophysical Large Area Survey''. Z.-Y.C. is supported by the China Postdoctoral Science Foundation grant No. 2014M560515.

% \doubleclearpage

%%%%%%%%%%%%%
% BIBLIOGRAPHY %
%%%%%%%%%%%%%

\end{document}